\documentclass[12pt]{article}
\pdfoutput=1
\usepackage{epsf,amsfonts,amssymb,epsfig,amsmath,mathtools,graphics,slashed}
\usepackage{hyperref,hep,graphicx,subfig}

\makeatletter

\newcommand{\Rmnum}[1]{\expandafter\@slowromancap\romannumeral #1@}
\makeatother

\newcommand{\nn}{\notag \\}

\begin{document}

\makeatletter
\renewcommand{\theequation}{\thesection.\arabic{equation}}
\@addtoreset{equation}{section}
\makeatother

\baselineskip 18pt

\begin{titlepage}

\vfill

\begin{flushright}
ICCUB-16-xxx\\
DCPT-16/21
\end{flushright}

\vfill

\begin{center}
   \baselineskip=16pt
   {\Large\bf Holographic Magnetisation Density Waves}
  \vskip 2cm
      Aristomenis Donos$^{1}$ and Christiana Pantelidou$^{2}$\\
   \vskip .6cm
      \vskip .6cm
      \begin{small}
      \textit{ $^{1}$ Centre for Particle Theory and Department of Mathematical Sciences,\\ Durham University
Durham, DH1 3LE, U.K \vspace{1cm}\\
      $^{2}$ Departament de Fisica Quantica i Astrofisica \& Institut de Ciencies del Cosmos (ICC)\\ Universitat de Barcelona, Marti i Franques 1, 08028 Barcelona, Spain.}
        \end{small}\\*[.6cm]

\end{center}

\vfill

\begin{center}
\textbf{Abstract}
\end{center}

\begin{quote}
We numerically construct asymptotically $AdS$ black brane solutions of $D=4$ Einstein theory coupled to a scalar and two $U(1)$ gauge fields. The solutions are holographically dual to $d=3$ CFTs in a constant external magnetic field along one of the $U(1)$'s. Below a critical temperature the system's magnetisation density becomes inhomogeneous, leading to spontaneous formation of current density waves. We find that the transition can be of second order and that the solutions which minimise the free energy locally in the parameter space of solutions have averaged stressed tensor of a perfect fluid.
\end{quote}

\vfill

\end{titlepage}
\setcounter{equation}{0}

\section{Introduction}

Phases that spontaneously break translational invariance are collectively referred to as Spatially Modulated (SM) phases and are widespread in Nature \cite{Seul476}. In condensed matter systems \cite{2009AdPhy..58..699V} in particular, they manifest themselves in various forms with spin density waves (SDW), and charge density waves (CDW) being the most common ones. Apart from the richness of this class of phases, its potential correlation with the physics of the pseudogap region of the high-$T_c$ superconductors' phase diagram \cite{Sachdev:2007qmc} and the conjecture that the QCD phase diagram at finite temperature and intermediate density is dominated by a chiral-density wave state\cite{CDW} have triggered a lot of interest in understanding these phases. 

Holography provides a natural theoretical framework to explore the physics of these phases at strong coupling. They are associated to black hole solutions that asymptote to AdS and have spatially modulated horizons. Many examples of modulated holographic phases have been discussed in the literature, starting with \cite{Donos:2012gg}  and further explored in \cite{Rozali:2012es,Donos:2013wia,Withers:2013loa,Withers:2013kva,Donos:2013woa,Donos:2012wi} (for earlier related work see \cite{Domokos:2007kt,Nakamura:2009tf,Ooguri:2010kt}).  It should be pointed out that constructing inhomogeneous black objects is not a new concept, dating back to the non-uniform strings. Gregory and Laflamme  showed that the uniform black string, described by the product of a Schwarzschild solution and a circle, becomes unstable to modes with wavenumber smaller than a critical value. Since the critical deformation mode is static, they pointed out a new family of black strings without translation invariance: the non-uniform strings \cite{Gregory:1994bj,Gregory:1993vy,Gregory:1987nb}. This new branch of solutions was perturbatively constructed \cite{Gubser:2001ac} and non-perturbatively in \cite{Wiseman:2002zc}. However, it was shown that the non-uniform solutions have higher mass than the uniform strings and thus they are not thermodynamically preferred in a sufficiently low number of dimensions \cite{Sorkin:2006wp}.

The electrically charged AdS-RN black branes have been shown to suffer from spatially modulated instabilities. Such instabilities have been investigated for a class of $D=5$ gravity theories with a single gauge-field and a Chern-Simons coupling in \cite{Nakamura:2009tf,Ooguri:2010kt,Donos:2012wi} and also for $D=4$ Einstein-Maxwell theory coupled to  a neutral pseudo-scalar field $\phi$ and possibly an additional gauge field in \cite{Donos:2011bh}.  The non-linear striped black branes, connected to the zero modes of \cite{Donos:2011bh}, have been constructed and studied in a series of papers \cite{Rozali:2012es,Donos:2013wia,Withers:2013loa,Withers:2013kva}. It was shown that, depending on the non-linearities of the model, a second order phase transitions occurs at some critical temperature $T_c$. Inhomogeneous solutions that break translation invariance in two directions were studied in \cite{Withers:2014sja},  where checkerboard black holes with rectangular fundamental domain were constructed. In the same work  \cite{Withers:2014sja}, it was also shown that the striped black holes are continuously connected to the checkerboard black holes via rectangular lattice black holes. In \cite{Donos:2012yu}, the electric stripes phase was shown to survive the existence of non-vanishing external magnetic, $B$.  The full three-dimensional family of solutions corresponding to oblique lattices was discussed in \cite{Donos:2015eew} and it was concluded that the free energy is minimised by a triangular lattice for $B$ greater than a critical value $B_{c}$. It was argued that this could be associated with minimal packing of circles in the plane.

Spatially modulated phases  of CFTs placed in magnetic field, in the absence of a chemical potential, have been investigated in \cite{Donos:2011qt}. A big class of $D$ dimensional bulk theories coupled to a scalar field, $\phi$ and to one or two $U(1)$ gauge fields was studied. It was shown that, for particular choices of Lagrangian parameters, the dual field theory can admit phases that spontaneously break translation invariance via current density waves. In these models, the key feature of the instability is that the relevant mode preserves the internal $U(1)$ symmetries. The zero modes that appear at the onset of the instability have been constructed in \cite{Donos:2011qt} and as we later explain, they modulate the magnetisation density of the dual field theory. Furthermore, it was shown \cite{Donos:2011pn} that these instabilities exist around magnetic branes solutions of $N = 8$ gauged supergravity in both $D=4$ and $D=5$. A different class of instabilities has been considered in e.g. \cite{Ammon:2011je,Bao:2013fda,Donos:2011pn} involving charged degrees of freedom, reminiscent of the instabilities discussed in \cite{Chernodub:2010qx}. In these case, the internal $U(1)$ under which the unstable fields are charged is spontaneously broken and the new phase will necessarily break translation.

Here, we construct the backreacted geometries dual to the magnetisation density waves found in \cite{Donos:2011qt}. This involves the numerical solution of a set of coupled partial differential equations in two coordinates, employing the DeTurck trick for dynamical gauge fixing  \cite{Headrick:2009pv} which was first used in a holographic set-up in \cite{Figueras:2011va}. For the specific model we examined, we find a branch of spatially modulated solutions that extends to lower temperatures and dominates the thermodynamic ensemble within our assumption of breaking translations in only one direction. For this branch,  we construct the complete two-dimensional space of solutions, specified essentially by the temperature  $T$ and the periodicity scale $k$. By minimising the free energy density, we determine the one dimensional branch of preferred solutions $k=k(T)$. We show that the latter have the stress-energy tensor of a perfect fluid. Down to the temperatures we explored, this branch seems to flow to a spatially modulated ground state with finite entropy. This is an interesting possibility which we plan to further explore in the future.

From the field theory point of view, we will examine a particle-hole symmetric medium which is charged under two different $U(1)$'s. The medium is magnetised with respect to one of them, under the influence of a uniform magnetic field. At temperatures below a critical value $T_{c}$, we will see that the magnetic densities become modulated giving rise to spontaneous magnetisation current densities. While the connection with holography is not direct, the magnetisation density is a natural variable to discuss SDWs in the context of hydrodynamics \cite{PhysRev.188.898}. One obvious difference with real systems exhibiting SDWs is the fact that the system we are considering is diamagnetic in its normal phase. Paramagnetic phases of holographic matter are certainly possible \cite{Donos:2012yu} and we expect that in a bottom-up approach similar phases can be constructed.

The remaining of this paper is organised as follows. In section \ref{sec2} we introduce the model of interest and review the spatially modulated instabilities of the magnetically charged AdS-RN black holes of \cite{Donos:2011qt}. In section \ref{sec3} we describe in detail the numerical method used to construct the back-reacted solutions and we report on the findings of this computation. Finally, we conclude in section \ref{sec4}. We have also included two Appendices were we discuss in some detail the asymptotic expansion of our solutions and the numerical convergence of the method we used to construct them.

\section{The setup}\label{sec2}

We consider a four dimensional theory of gravity coupled to 
a scalar field, $\phi$, and two gauge fields, $A$ and $B$ gauging the corresponding symmetries $U(1)_{A}$ and $U(1)_{B}$ in the bulk. The bulk dynamics will be described by the Lagrangian
\begin{align}\label{eq:Lagra}
\mathcal{L}= &\tfrac{1}{2}R-V\left(\phi\right)-\tfrac{1}{2} (\partial\phi)^{2}-\tfrac{1}{4}Z_{A} \left(\phi\right)\,F_{\mu\nu} F^{\mu\nu}\nn&-\tfrac{1}{4} Z_{B}(\phi) G_{\mu\nu}G^{\mu\nu}-\frac{1}{2}W\left(\phi\right)\,F_{\mu\nu}G^{\mu\nu}\, 
\end{align}
where $F= dA$ and  $G= dB$. The above form is natural from the point of view of $N=2$ supergravity in $D=4$ \cite{Andrianopoli:1996cm,Andrianopoli:1996vr} which commonly appears in the context of SUSY consistent truncations of $D=11$ SUGRA (see e.g. \cite{Cvetic:1999xp,Gauntlett:2009zw,Donos:2010ax,Bobev:2010ib}). Note that the wedge product terms between the field strengths that appear in top-down models are not going to be an important structural difference since they would be inactive for the solutions we consider in this paper. The equations of motion deriving from \eqref{eq:Lagra} are given by
\begin{align}\label{eoms}
&R_{\mu\nu}=V\,g_{\mu\nu}+\partial_{\mu}\phi\,\partial_{\nu}\phi
+Z_{A} \left(F_{\mu\rho}F_{\nu}{}^{\rho}-\tfrac{1}{4}g_{\mu\nu}F_{\rho\sigma}F^{\rho\sigma} \right)\nn
&\qquad\quad+ Z_{B}\left(G_{\mu\rho}G_{\nu}{}^{\rho}-\tfrac{1}{4}g_{\mu\nu}G_{\rho\sigma}G^{\rho\sigma}\right)
+2W \left( G_{(\mu}{}^\rho F_{\nu)\rho}{}-\tfrac{1}{4}g_{\mu\nu}G_{\rho\sigma}F^{\rho\sigma} \right)\, ,\nn
&\nabla_{\mu}\left(Z_{A}\, F^{\mu\nu}+W\,G^{\mu\nu}\right)=0\, ,\nn
&\nabla_{\mu}\left(Z_{B}\,G^{\mu\nu}+W\,F^{\mu\nu} \right)=0\, ,\nn
&\nabla^{2} \phi-V^{\prime}-\tfrac{1}{4}Z_{A}^{\prime}\,F_{\mu\nu} F^{\mu\nu}
-\tfrac{1}{4}Z_{B}^{\prime}\,G_{\mu\nu}G^{\mu\nu}-\tfrac{1}{2} W^\prime\,G_{\mu\nu}F^{\mu\nu}=0\, .
\end{align}
The functions $V,Z_{A},Z_{B}$ and $W$ are chosen to admit the following expansion
\begin{align}\label{conds}
V(\phi)&=-6+\tfrac{1}{2}\,m_{s}^{2}\,\phi^{2}+\cdots,\nn
Z_{A}(\phi)&=1-n\,\phi^{2}+\cdots,\nn
Z_{B}(\phi)&=1+\cdots , \nn
W(\phi)&=s\,\phi+\cdots\, .
\end{align}
around $\phi=0$, where $m_s,n$ and $s$ are constants. The equations of motion \eqref{eoms} admit $AdS_4$ of radius $R^{2}_{AdS_{2}}=1/2$ as a solution and will serve as the asymptotics of all the black hole space times we are going to construct.

According to the scenario of the introduction, we now consider the deformation of the boundary theory by a magnetic field $\beta$ in the $U(1)_{A}$. At high temperatures the bulk geometry is going to be described by the magnetic AdS-RN black hole,
\begin{align}\label{eq:RNsol}
ds^{2}&=\frac{1}{z_{h}^{2}\,z^2}\left(-f\,dt^{2}+\frac{z_{h}^{2}}{f}\,dz^{2}+L^{2}\,dx^{2}+dy^{2}\right)\, ,\nn
F&=L\,\beta\,dx\wedge dy\,,\nn
f&=z^3\left(\frac{2}{z^3}+z\frac{\beta^2 z_{h}^{4}}{2}\right)-z^3\left(2+\frac{\beta^2 z_{h}^{4}}{2}\right)
\end{align}
and $\phi$, $B$ trivial. This is a solution allowed by the choice \eqref{conds} which lets us set the scalar $\phi$ and the second gauge field $B$ consistently to zero. We have also introduced a length-scale $L$ for later convenience when we fix the period of $x$. In this coordinate system the conformal boundary is at $z=0$ with
\begin{align}\label{eq:boundary_metric}
ds_{4}^{2}=\frac{d\varepsilon ^{2}}{2\,\varepsilon^{2}}+\frac{1}{\varepsilon^{2}}\,(-2\,dt^{2}+L^{2}\,dx^{2}+dy^{2})\,.
\end{align}
where we defined $\varepsilon=z\,z_{h}$. The horizon of the black hole \eqref{eq:RNsol} is at $z=1$ and its Hawking temperature is $T=\frac{12-\beta^2\, z_h^4}{8\pi\, z_h}$. The near horizon limit of the extremal solution with $T=0$ reduces to
\begin{align}
\label{eq:AdS_solution}
ds^{2}&=\frac{1}{12}ds^2\left(AdS_{2}\right)+d\hat{x}^{2}+d\hat{y}^{2},\nn
F&=\sqrt{12}\,d\hat{x}\wedge d\hat{y}\, ,
\end{align}
where $ds^{2}\left(AdS_{2}\right)$ is the metric on a unit radius $AdS_{2}$ space and we have scaled the spatial field theory coordinates according to e.g. $\hat{x}=L\,(\beta/12)^{1/2}\,x$. 
 
We now turn our attention to the instabilities of the magnetic AdS-RN black hole \eqref{eq:RNsol} towards phases with broken translations that were discussed in \cite{Donos:2011qt}. These can be understood as near horizon instabilities by examining the perturbation
\begin{align}\label{fbpert}
\phi=\delta\phi\left(x^{\alpha}\right)\,\cos\left(k\,\hat{x}\right),\qquad 
B=\delta B\left(x^{\alpha}\right)\,\sin\left(k\,\hat{x}\right)\,dy\, ,
\end{align}
around the background solution \eqref{eq:AdS_solution}. The functions that appear in \eqref{fbpert} depend on the coordinates of $AdS_{2}$ and  
$k$ is a constant. The linearised equations of motion \eqref{eoms} around \eqref{eq:AdS_solution} take the form
\begin{align}
\Box_{2}{\bf v}-\tfrac{1}{12} M^2{\bf v}&=0\, ,
\end{align}
where  ${\bf v}=(\delta\phi,\delta B)$ and the Laplacian is with respect to the $AdS_{2}$ metric. The mass matrix is given by 
\begin{align}\label{eq:mass_matrix}
M^{2}=\left(
   \begin{matrix} 
      \tilde{m}_{s}^{2}+k^{2} & 2 \sqrt{3}\, s\,k \\
     2\sqrt{3}\, s\,k & k^{2} \\
   \end{matrix}
   \right)\, ,
\end{align}
with $\tilde{m}_{s}^{2}\equiv m_{s}^{2}-12 n$.
It can be easily checked that depending on the choices of  $n$ and $s$, the eigenvalues of the above matrix can violate the BF bound signalling an instability. Interestingly, the lightest mode appears generically at a finite value of $k$. 

\begin{figure}[h]
\centering
\subfloat{\includegraphics[height=5.5cm]{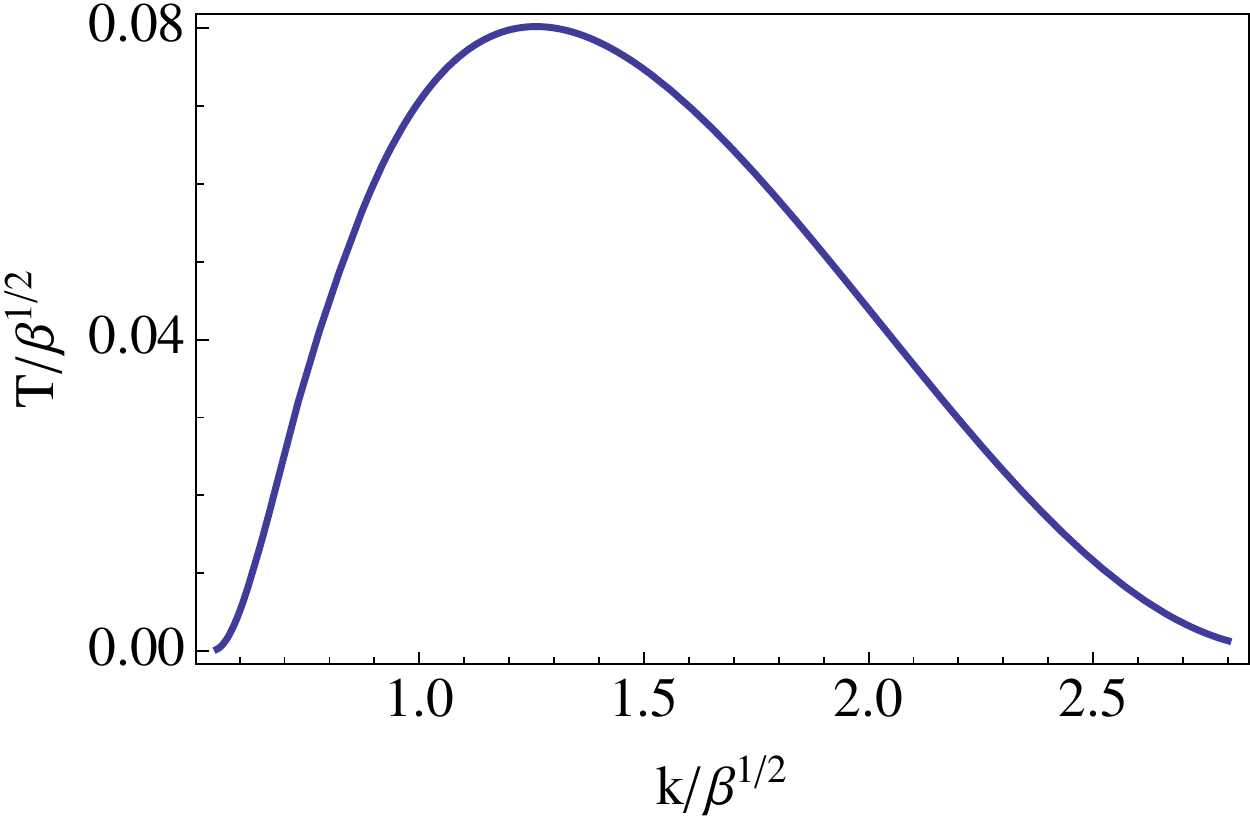}\label{fig:a}}
\caption{Plots of critical temperatures $T$ versus $k$ for the existence of normalisable zero modes
about the $D=4$ magnetically charged
AdS-RN black hole solutions with $n=-1$ and $s= 2$.}
\label{fig:ZeroModes}
\end{figure}

The finite temperature zero modes related to the above instabilities have been constructed in \cite{Donos:2011qt} for $m_s^2=-4$ and various values of $n,s$. These were constructed by examining perturbations around the geometry of the black holes \eqref{eq:RNsol}. The operator $\mathcal{O}_\phi$, dual to the scalar field $\phi$, was chosen to have scaling dimension $\Delta=1$ and it was shown that the zero modes appear along a curve $T(k)$, specified by the parameters $n$ and $s$. At each point on that curve, a new branch of broken phase black hole solutions is expected appear for the corresponding fixed value of $k$. The maximum of this curve reveals the critical temperature in the case of a continuous transition. In figure \ref{fig:ZeroModes} we plot $T(k)$ for the case 
$(n,s)=(-1,2)$, which corresponds to $(T_c,k_c)=\beta^{1/2}\,(0.08,1.26)$. 

The appearance of the bulk gauge field $B$ in the mode \eqref{fbpert} and the fact the instability shows up at a finite value of $k$, suggest that the broken phase will develop an inhomogeneous current density. As we explain in section \ref{sec:Thermodynamics}, these currents can be seen as magnetisation currents due to the modulation of the field theory magnetisation density. At the same time, the operator dual to the bulk scalar $\phi$ will take a modulated VEV with the remaining fields backreacting at second order in perturbation theory close to the critical temperature $T_{c}$.

In this note, we go beyond perturbation theory constructing the backreacted geometries by solving the non-linear equations of motion \eqref{eoms}. As expected, the full functional dependence of $V, Z_{A}, Z_{B}$ and $W$ on $\phi$ is going to be relevant. The model we are going to consider has
\begin{align}\label{conds2}
V(\phi)&=-6-2\phi^{2}+\phi^{4},\nn
Z_{A}(\phi)&=1+\,\phi^{2}\,,\nn
Z_{B}(\phi)&=1\, , \nn
W(\phi)&=2\,(\phi-\phi^{3})\,.
\end{align}

The choice \eqref{conds2} has $m_s^2=-4$, $n=-1$ and $s=2$. The zero modes appearing in this model correspond to the bell curve shown in figure \ref{fig:ZeroModes}. Furthermore, note that the model \eqref{conds2} exhibits a $\mathbb{Z}_2$ symmetry with $b_y\to -b_y$ and $\phi\to-\phi$ simultaneously. This discrete symmetry becomes important at the non-linear level, relating the two branches of solutions expected to emerge at the critical temperature $T_c$. 
If the $\mathbb{Z}_2$ is not present, the two branches will be distinct and they will therefore have different thermodynamic properties.

 \section{The broken phase black holes}\label{sec3}
 
 In this section we discuss the construction and properties of the non-linear solutions corresponding to the new branches of black holes proposed to exist in \cite{Donos:2011qt} and reviewed in section \ref{sec2}. In section \ref{sec:setup} we describe the boundary value problem relevant to the present physical situation and the numerical methods we used to solve it. In section \ref{sec:Thermodynamics} we discuss the thermodynamics as well as the local magnetisation properties of the field theory states dual to the newly constructed black holes. In section \ref{sec:num_sols} we discuss the numerical solutions we constructed along with some of their properties.
 
 \subsection{The ansatz and the method used}
 \label{sec:setup}
 We consider the following ansatz for the back-reacted solutions
  \begin{align}
 \label{eq:ansatz}
 &ds^2=\frac{1}{z_{h}^{2}\,z^{2}}\left[-f\,Q_{tt}\,dt^2+\frac{z_{h}^{2}\,Q_{zz}}{f}\,dz^2+Q_{xx}\,( L\,dx+z^2Q_{zx}dz)^2+Q_{yy}\,dy^2\right]\,\nonumber\\
 &A=(\beta L\, x+a_y )dy\,\nonumber\\
 &B=b_y dy\,\nonumber\\
 &\phi= z\, h\,,
 \end{align}
 where $Q_{tt},Q_{zz},Q_{xx},Q_{yy},Q_{zx},a_y,b_y$ and $h$ are functions of the radial coordinate, $z$, and $x$. Furthermore, we require that these functions obey periodic boundary conditions in the $x$ direction, with period given by $L$. The function $f(z)$ is the same with the one in equation \eqref{eq:RNsol}. Note that this ansatz is generic enough to capture both the normal phase solution \eqref{eq:RNsol}, corresponding to $Q_{tt}=Q_{zz}=Q_{xx}=Q_{yy}=1$, $Q_{zx}=Q_{tx}=h=a_y=b_y=0$, as well as the static zero modes considered at linearised level in \cite{Donos:2011qt}.  In terms of the functions appearing in the non-linear ansatz \eqref{eq:ansatz}, this mode takes the form
\begin{align}\label{eq:bh_mode}
\delta h(z,x)&=H(z)\,\cos(2\pi x)\notag\\
\delta b_{y}(z,x)&=B_{y}(z)\,\sin(2\pi x)\,.
\end{align}
    
The PDEs obtained when the equations of motions are evaluated on this ansatz are weakly elliptic, meaning that they are elliptic only for the physical degrees of freedom, and thus are unsuitable for numerics without gauge fixing. In this work, we employ the DeTurck method to resolve this issue \cite{Headrick:2009pv,Figueras:2011va}.  According to this method, instead of solving the Einstein equations, one solves the Einstein-DeTurck equations  which are obtained from \eqref{eoms} after making the following shift
 \begin{equation}
 R_{\mu\nu}\to R_{\mu\nu}+\nabla_{\mu}\xi_{\nu}
 \end{equation}
 where $\xi^\mu=g^{\nu\lambda}(\Gamma^\mu_{\nu\lambda}(g)-\bar{\Gamma}^\mu_{\nu\lambda}(\bar{g}))$. Here $\bar{g}$ denotes a reference metric (which is required to have the same asymptotic behaviour as $g$) and $\bar{\Gamma}$ is the Christoffel connection of $\bar{g}$; we choose this to be the AdS-RN black hole metric \eqref{eq:RNsol}. The resulting PDEs are then strictly elliptic and, with appropriate boundary conditions, can be solved numerically using a relaxation method. 

In more detail, one discretises the coordinates of the PDEs to form a lattice. Our coordinates span $z\,\epsilon[0,1]$ and $x\,\epsilon[0,1)$. To approximate the derivatives of our functions at each grid point, we use a Fourier expansion in the periodic direction $x$. The treatment of the $z$ direction is a more delicate one since our functions are non-analytic close to the boundary of $AdS_{4}$ at $z=0$ as we explain in Appendix \ref{app:boundary_conditions}. We have used both spectral methods on a Chebyshev collocation grid as well as a fourth order finite difference scheme to cross check our results finding essentially the same outcomes within our numerical precision. We have included a convergence test in appendix \ref{app:Numerical} showing power law convergence for the spectral method in the $z$ direction with a power compatible with the non-analytic terms in our expansion.

After fixing a discretisation scheme, the problem then reduces to solving a set of non-linear algebraic equations for the values of our functions on the grid described above. This is done using the Newton-Raphson method where one starts with an initial guess for the unknown functions at each lattice point, which presumably does not solve the PDEs. The solution is then iteratively improved using Newton's method in order to obtain functions that solve the PDEs to a better and better approximation. This procedure leads to a countable set of (locally unique) solutions to the modified equations of motions. A question that naturally arises is whether these solutions are also solutions of the initial equations \eqref{eoms}. This is true when $\xi^2=0$\footnote{Since $\xi^\mu$ is spacelike, $\xi^\mu=0$ is equivalent to $\xi^2=0$.}, i.e. when solutions corresponding to Ricci solitons are discarded. For this reason, after generating the solutions, we check that $\xi^2$ is zero within numerical precision at all grid points. 

As we explain in more detail in Appendix \ref{app:boundary_conditions}, a set boundary conditions at $z=0$ which are appropriate for $AdS_{4}$ asymptotics are
 \begin{align}
 &Q_{tt}(0,x)=Q_{xx}(0,x)=Q_{yy}(0,x)=1\,\nonumber\\
 &Q_{zx}(0,x)=a_y(0,x)=b_y(0,x)=\frac{\partial h(0,x)}{\partial z}=0\,.
 \end{align}
On the other end of our computational domain, located at $z=1$ we need to impose boundary conditions which guarantee a smooth Killing horizon of temperature $T=\frac{12-\beta^2\, z_h^4}{8\pi\, z_h}$. This boils down to demanding that the functions, $\mathcal{F}(z,x)$ that parametrise our ansatz \eqref{eq:ansatz}, admit an analytic expansion of the form
 \begin{align}
 \label{IRexp}
 \mathcal{F}=\mathcal{F}(1,x)-(1-z)\partial_z\mathcal{F}+\dots\,.
 \end{align}
The equations of motion impose constraints on the coefficients of the power series \eqref{IRexp}. In order for the Euclidean signature metric to have a smooth fixed point at $z=1$ we must have $Q_{zz}(1,x)=Q_{tt}(1,x)$. By expanding the equations of motion at $z=1$ we find another seven relations describing Robin boundary conditions imposed on that surface. Throughout the calculation, the magnetic field will be set to $\beta=1$ while $z_h$ will be tuning the temperature.
 
A final point we need to address is the Goldstone mode associated with the spontaneous nature of the way we break translations. More concretely, if $\mathcal{F}(z,x)$ is a solution, then $\mathcal{F}(z,x+c)$ is also a solution of the boundary value problem for any constant $c$. Because of this, we should impose a condition in order to mod out these solutions from the moduli space. It is enough to impose this condition at one point and for the specific solutions we constructed, we imposed $b_y(1,0)=0$.

\subsection{Thermodynamics and magnetisation currents}\label{sec:Thermodynamics}
To analyse the thermodynamics of our black hole solutions, we need to regularise our bulk action by adding appropriate surface terms \cite{Balasubramanian:1999re,Bianchi:2001kw}. For our bulk theory \eqref{eq:Lagra}, this can be achieved by adding a surface term $S_{\partial}$ to the bulk action
\begin{equation}
S_{reg}=S+S_{\partial}\,,
\end{equation}
which includes both the Gibbons-Hawking term as well as the necessary counterterms that render the total action finite. More specifically\footnote{The specific choice of scalar field counterterms is certainly compatible with the $\Delta=1$ choice for the boundary operator. However, when supersymmetry is involved a more careful treatment is required as pointed out in \cite{Freedman:2013ryh}.},
\begin{equation}\label{ctermp}
S_{\partial}=\int d\tau d^3 x \sqrt{-g_\infty}\,(K+\frac{1}{\sqrt{2}}(4-\phi^2)+\phi\,\eta^\mu\partial_\mu \phi+ \cdots)\,.
\end{equation}
Here $K=g^{\mu \nu} \nabla_\mu n_\nu$ is the trace of the extrinsic curvature of a $z=const$ surface close to the boundary, with $n^{\mu}$ the dual of the outward pointing normal unit vector and $g_\infty$ being the determinant of the projection of the bulk metric on that surface. The ellipsis refers to terms which will not be relevant for the ansatz and boundary conditions that we are considering. 

Following \cite{Balasubramanian:1999re,Bianchi:2001kw}, we now compute the expectation value of the boundary stress-energy tensor. 
The relevant terms are given by 
\begin{equation}\label{stressy}
\langle T^{\mu\nu}\rangle=\lim_{z \to 0}\,\frac{1}{(z\,z_{h})^{5}}\, [K^{\mu\nu}-K\,g_{\infty}^{\mu\nu}-\frac{1}{\sqrt{2}}\left(4-\phi^2\right)g_{\infty}^{\mu\nu}-\phi\,\eta^\rho\partial_\rho \phi \,g_{\infty}^{\mu\nu}+\cdots]\,,
\end{equation}
where we have included an appropriate power of $z_{h}$ since we are interested in the CFT one-point functions with respect to the metric
\begin{align*}
ds_{3}^{2}=\gamma_{\mu\nu}\,dx^{\mu}\,dx^{\nu}=-2\,dt^{2}+L^{2}\,dx^{2}+dy^{2}\,.
\end{align*}
In order to express this in terms of our asympotic data, we need to plug in the expansion of our fields close to the $z=0$ boundary. Using the expansion \eqref{UVexp} given in Appendix \ref{app:boundary_conditions}, we obtain
\begin{align}\label{stress}
&\langle T^{t}{}_{t}\rangle=-\frac{1}{2\,\sqrt{2}\, z_h^3}(4+\beta^2\, z_h^4-6Q_{tt}^{(3)})\,,\nonumber\\
&\langle T^{x}{}_{x}\rangle=\frac{1}{4\sqrt{2}\, z_h^3}(4+\beta^2\, z_h^4+12Q_{xx}^{(3)})\notag\\
&\langle T^{y}{}_{y}\rangle=\frac{1}{4\sqrt{2}\, z_h^3}(4+\beta^2\, z_h^4+12Q_{yy}^{(3)})\,.
\end{align}
and as we discuss in Appendix \ref{app:boundary_conditions}, the equations of motion imply that $\langle T_{\mu\nu}\rangle$ is traceless. Defining the unit-norm timelike vector $u=2^{-1/2}\,\partial_t$ on the boundary, we can express the free energy $\bar{w}$, mass $\bar{m}$ and entropy $\bar{S}$ averaged densities in terms of our numerical data
\begin{align}\label{eq:free_energy}
\bar{w}&=\bar{m}-T\,\bar{S}\notag\\
\bar{m}&=L^{-1}\,\int_{0}^{1}\sqrt{-\gamma}\,u^\mu u^\nu\langle T_{\mu\nu}\rangle dx\notag\\
\bar{S}&=2\pi\,z_h^{-2}\, \int_0^{1} \,dx\,\sqrt{Q_{xx}(1,x)\,Q_{yy}(1,x)}\,.
\end{align}

We now turn our attention to the electric currents of the dual field theory. Varying the on-shell action with respect to the asymptotic value of the gauge fields we obtain their VEVs
\begin{align}\label{eq:b_currents}
 &\langle J_{A}^{\mu}\rangle=- \lim_{z \to 0}\,\frac{1}{(z\,z_{h})^{3}}\, n_{\rho}\,\lbrack Z_{A}\,F^{\rho\mu} +W\,G^{\rho \mu}\cdots\rbrack\notag\\
 &\langle J_{B}^{\mu}\rangle=- \lim_{z \to 0}\,\frac{1}{(z\,z_{h})^{3}}\, n_{\rho}\,\lbrack Z_{B}\,G^{\rho\mu} +W\,F^{\rho\mu}\cdots\rbrack\,.
 \end{align}
for which the expansion \eqref{UVexp} gives
\begin{equation}
\langle J_{A}^{y}\rangle=\sqrt{2}\,z_{h}^{-1}\, j_a\,,\quad \langle J_{B}^{y}\rangle=\sqrt{2}\,z_{h}^{-1}\, j_b\,,
 \end{equation}
for the non-zero components. As a comment, the only non-trivial component of the stress tensor continuity equation
\begin{equation}
Q_{x\,x}^{(3)}{}^{\prime}=\frac{2}{3} z_h^2\beta\,L \, j_a
\end{equation}
is satisfied provided that  $\xi^{\mu}=0$ on the background . This consists another check we performed on our numerics.

For a fixed period $L$, the first law of thermodynamics gives
\begin{equation}\label{eq:first_law}
\delta \bar w=-\bar{S}\, \delta T -\bar M_{A} \,\delta \beta\,,
\end{equation}
where $\bar{M}_{A}$ is the thermodynamic magnetisation corresponding to $U(1)_{A}$. For the backgrounds we construct in this paper, the magnetisation densities of both $U(1)$'s are inhomogeneous. Our black holes are bulk duals of equilibrium states and one would expect that the currents of the boundary theory should be a total derivative of a periodic antisymmetric rank two magnetisation tensor resulting in zero net transport of  charge. To show this from the bulk \cite{Donos:2015bxe}, we integrate the gauge field equations of motion \eqref{eoms} over the radial coordinate $z$. Using regularity on the horizon and the definition \eqref{eq:b_currents} we can express $\sqrt{-\gamma}\,\langle J^{i}_{A, B}\rangle=\partial_{j}M_{A,B}^{ij}$ with $M_{A, B}^{xy}=-M^{yx}_{A,B}=M_{A,B}$ and
\begin{align}\label{eq:l_magnetisation}
M_{A}&=-\int_{0}^{1} dz \sqrt{-g} (Z_{A}\, F^{x\,y}+W\, G^{x\,y})\notag\\
M_{B}&=-\int_{0}^{1} dz \sqrt{-g} (Z_{B}\, G^{x\,y}+W\, F^{x\,y})\,. 
\end{align}
Note that the definition of the magnetisation through the current density leaves us with a constant unfixed. We fix this constant by demanding that the thermodynamic magnetisation, defined through \eqref{eq:first_law}, is equal to the average of the local magnetisation. This is achieved by the choice we make when we write equation \eqref{eq:l_magnetisation}.

Following the arguments of \cite{Donos:2013cka}, we find that the variation of the average free energy density with respect to the period $L$ is given by
\begin{equation}
-L\frac{\delta \bar{w}}{\delta L}= \bar{w}+\beta\,\bar{M}_{A} +\bar{T}^x{}_x\,.
\end{equation}
Even though the solutions we construct numerically break translations in the $x$ direction only, an identical statement is true regarding the variation of the free energy with respect to the period of the $y$ direction. However, since our solutions don't depend on $y$, for all the solutions we construct here
\begin{align}
\bar{w}+\beta\,\bar{M}_{A}+\bar{T}^{y}{}_{y}=0\,.
\end{align}
From the above we conclude that for the black holes holes which locally minimise their free energy as a function of $L$, we have
\begin{align}
\bar{w}_{min}&=-\beta\,M_{A} -P,\quad \bar{T}^x{}_x=\bar{T}^y{}_y=P
\end{align}
and $T_{xy}=0$ by construction. For these solutions, we see that the averaged stress tensor corresponds to a perfect fluid.

Scaling invariance of the dual CFT constrains further the various thermodynamic quantities we have discussed. In particular, the free energy has to be expressible according to  $\bar{w}=\beta^{3/2}\,f(T/\sqrt{\beta},L\, \sqrt{\beta})$ with $f$ a dimensionless function. This symmetry therefore constrains the thermodynamic magnetisation according to
 \begin{align}\label{eq:M_scal_inv}
\beta\,\bar{M}_{A}=- \beta\,\frac{\partial \bar{w}}{\partial \beta}
=-\frac{3}{2}\,\bar{w}-\frac{1}{2}T\,\bar{S}+\frac{1}{2}\,L\frac{\delta \bar{w}}{\delta L}
 \end{align}
 Moreover, equations \eqref{eq:free_energy} and \eqref{eq:M_scal_inv} imply that $\bar{T}_{tt}=4P$ in agreement with the stronger requirement on the stress tensor being traceless locally.

For the normal phase black hole solution \eqref{eq:RNsol} we can simply evaluate the thermodynamic magnetisation and susceptibility \cite{Donos:2012yu} in terms of $T$ and $\beta$. Using \eqref{eq:l_magnetisation} we find
\begin{align}\label{eq:magnetisation_np}
M_{A}&=-\beta\,z_{h}(T,\beta),\quad M_{B}=0\\
\chi_{A}&\equiv\frac{\partial \bar{M}_{A}}{\partial\beta}=-\frac{1}{3}z_{h}\,\frac{12+\beta^{2}\,z_{h}^{4}}{4+\beta^{2}\,z_{h}^{4}}\,,
\end{align}
where the minus sign in the susceptibility points out that the normal phase describes a diamagnet with respect to $U(1)_{A}$.
It is easy to see from the leading mode \eqref{eq:bh_mode}, that close to $T_{c}$ the broken phase black holes will have a modulated $U(1)_{B}$ magnetisation density according to
\begin{align}
\delta M_{B}\approx M_{B}^{0}\,\left(1-T/T_{c}\right)^{1/2}\,\cos(2\pi x)
\end{align}
with higher order corrections away from $T_{c}$. The effects of inhomogeneity on the magntisation of $U(1)_{A}$ will be second order. We show numerical evidence for this behaviour in the next section where we discuss the backreacted solutions.

 \subsection{Numerical solutions}\label{sec:num_sols}
 
 \begin{figure}[h]
\centering
\subfloat{\includegraphics[height=5.5cm]{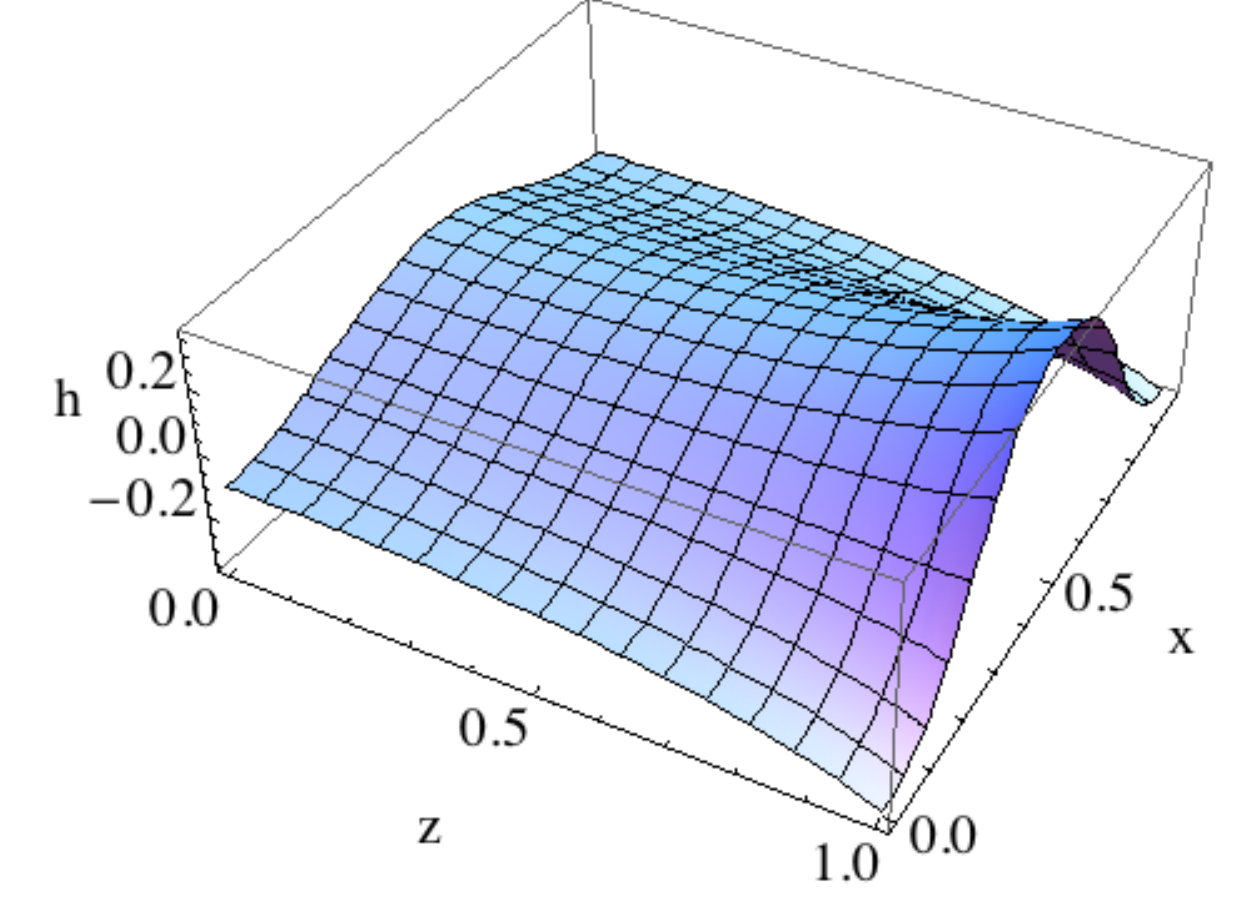}\quad\includegraphics[height=5.5cm]{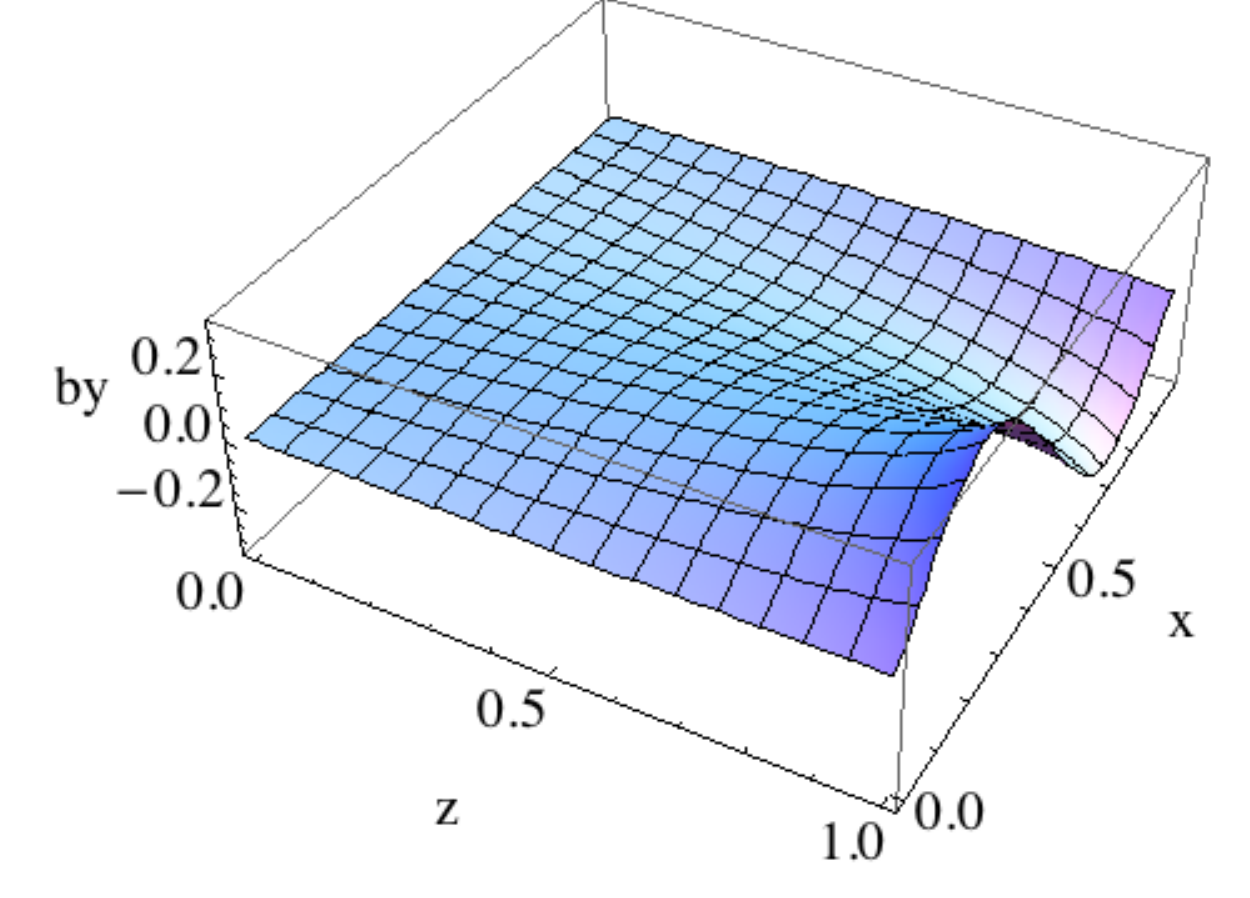}}
\caption{The scalar field, $h$, and the gauge field $b_y$ for $T = 0.5\, T_c$ and $k_c=k$. $N_x = 40, N_z = 40$.}
\label{fig:RepSln_05Tc}
\end{figure}

 In this section we present the solutions constructed for the model fixed by \eqref{conds2}. Due to its $\mathbb{Z}_2$ symmetry we expect to find only one distinct branch of solutions. From the numerics we see that this branch extends to lower temperatures, $T<T_c$, covering  the two-dimensional region below the bell-curve  $T(k)$ shown in figure \ref{fig:ZeroModes}.
We present the profiles of some of the functions in figure \ref{fig:RepSln_05Tc}, for a representative solution with $T = 0.5\, T_c$ and $k_c=k$.
 
\begin{figure}[h]
\centering
\subfloat{\includegraphics[height=7cm]{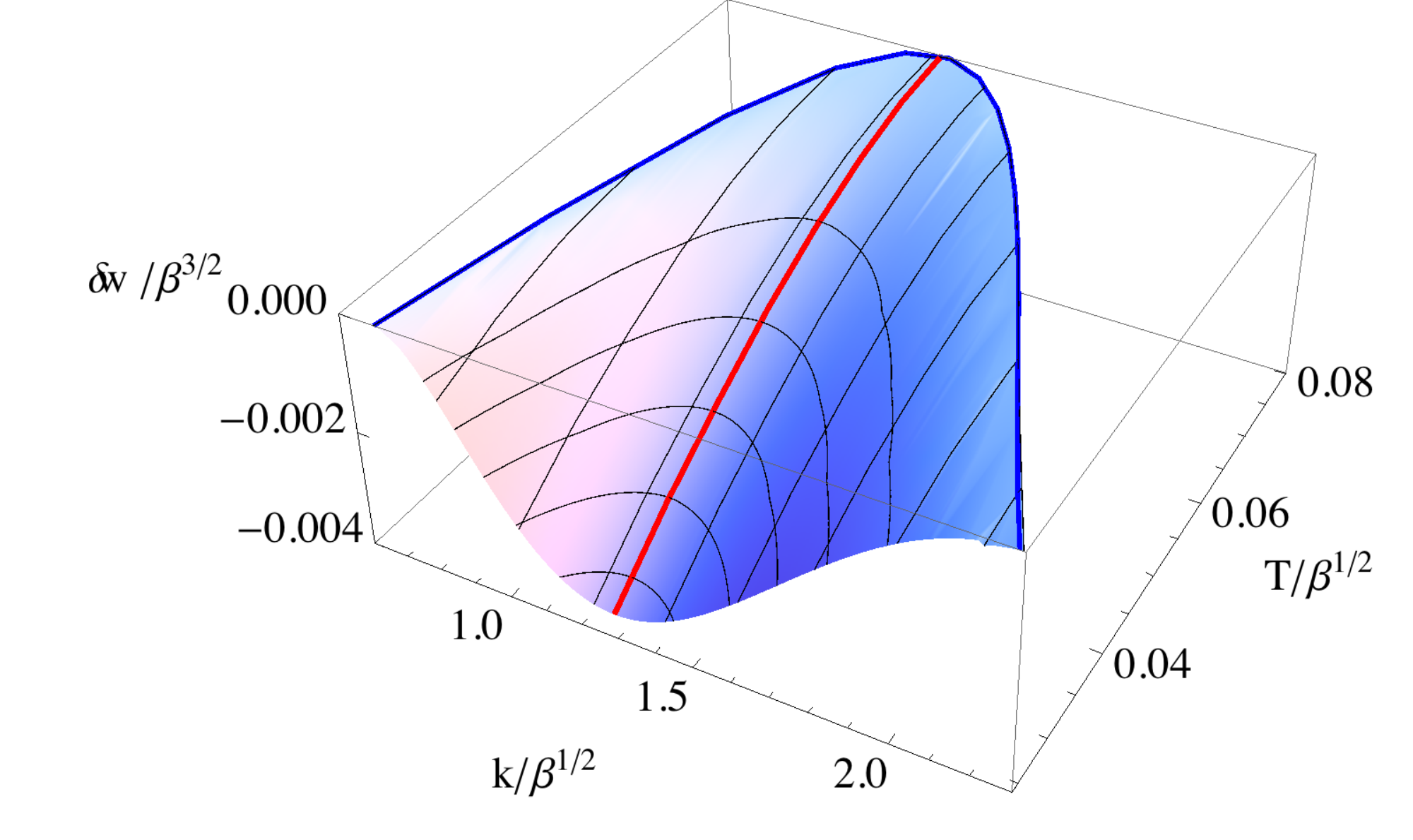}}
\caption{The difference between the free energy of the normal and the modulated phase for the whole moduli space of solutions}
\label{fig:Moduli}
\end{figure}

\begin{figure}[h]
\centering
\subfloat{\includegraphics[width=7cm]{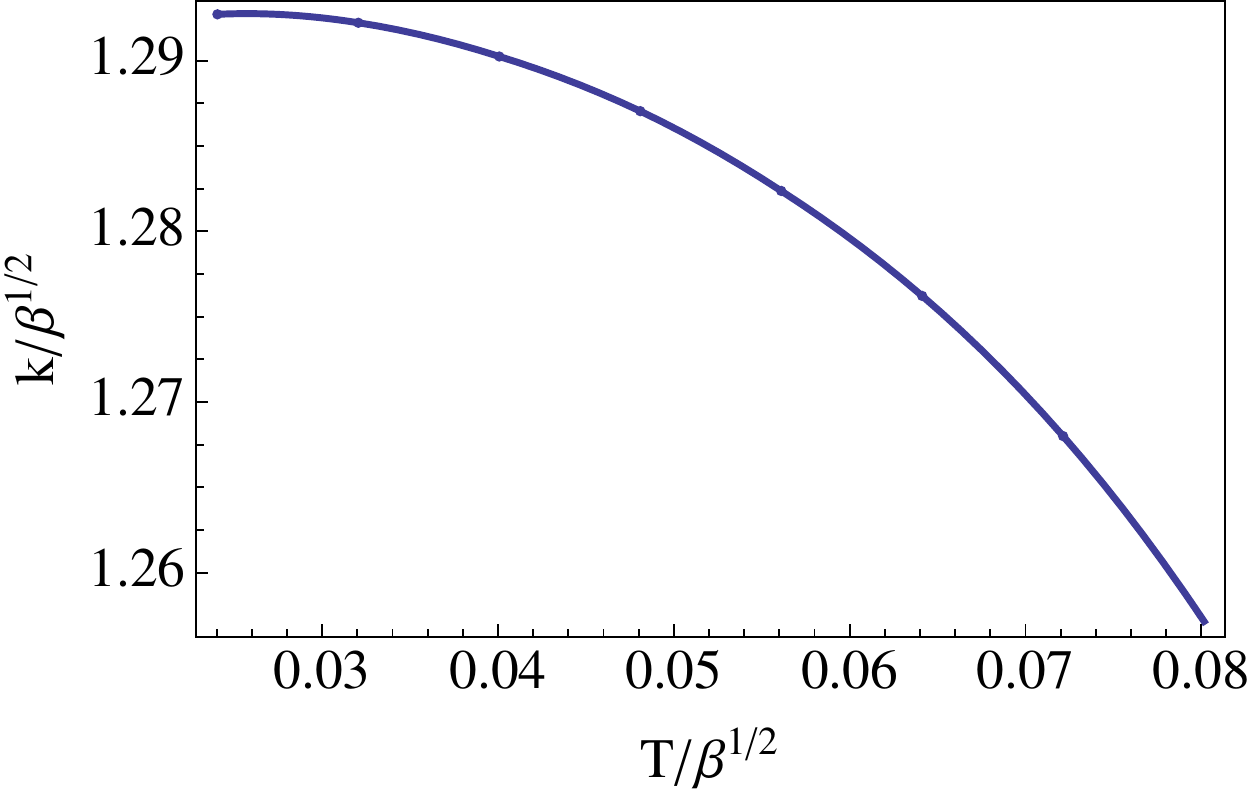}\quad\includegraphics[width=7cm]{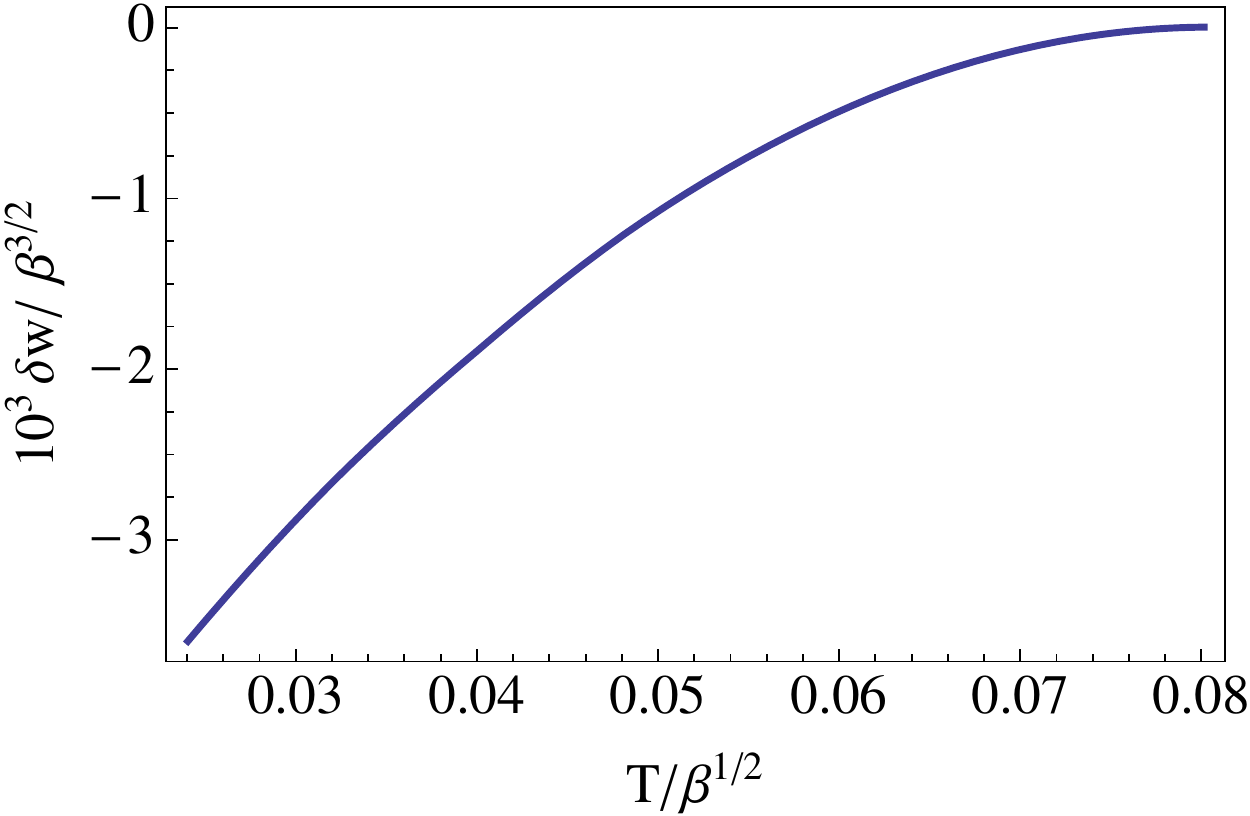}}\\\subfloat{\includegraphics[width=7cm]{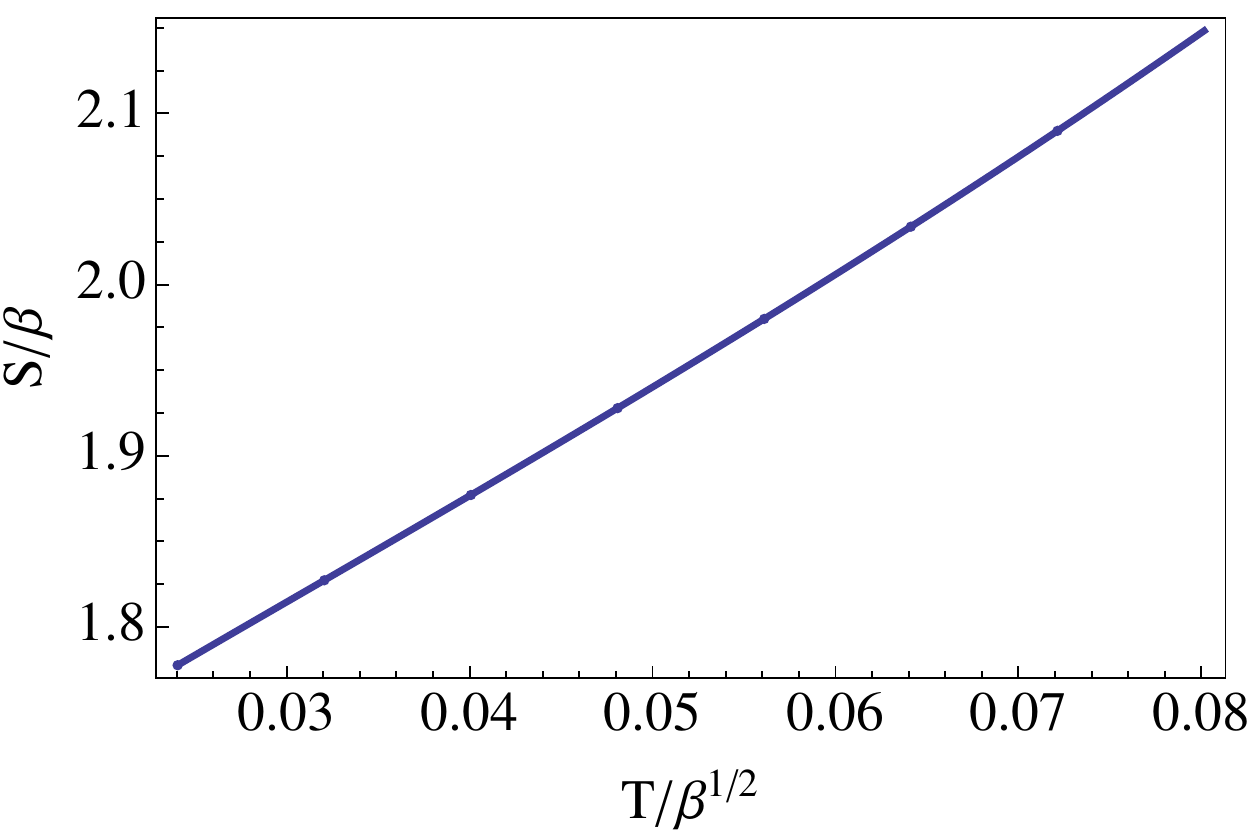}}
\caption{Properties of the preferred branch. }
\label{fig:PrefBranch}
\end{figure}

\begin{figure}[h]
\centering
\subfloat{\includegraphics[height=6cm]{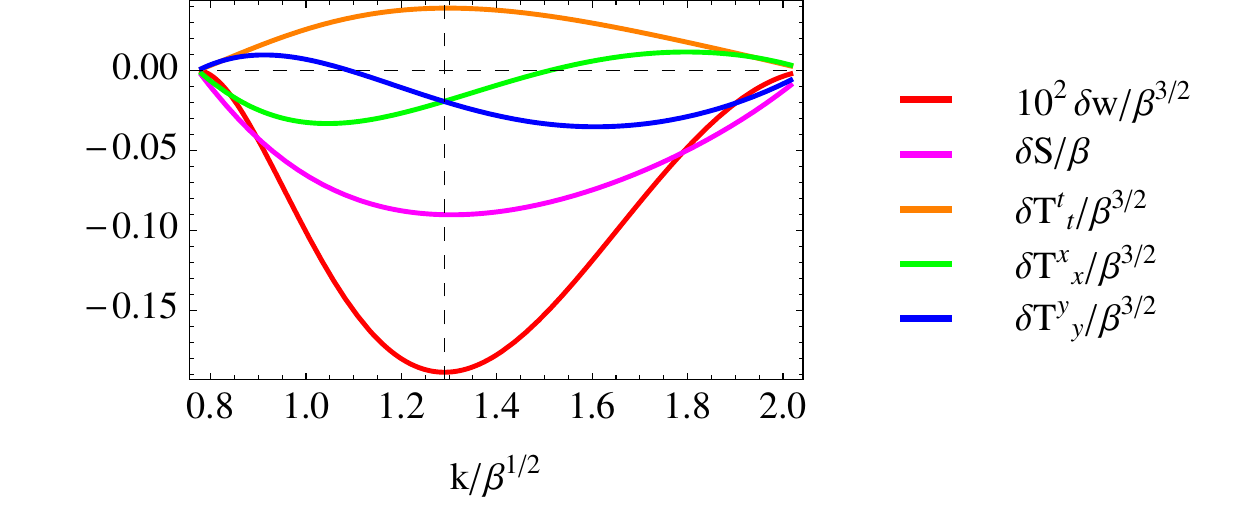}}
\caption{Free energy, entropy and averaged stress tensor for $T=0.5\, T_c$. $N_x = 40, N_z = 40$.}
\label{fig:Thermo_05Tc}
\end{figure}

  In figure \ref{fig:Moduli}, we show a 3D plot of the free energy difference between the normal and the modulated phase, $\delta w$,  as a function of $T$ and $k$ and we see that through out the moduli, the modulated solutions dominate.
 To specify the thermodynamically preferred branch branch of black holes one needs to minimise the free energy $w$ with respect to $k$ for fixed $T$. Doing this numerically yields the red line in figure \ref{fig:Moduli}. Various properties of the preferred branch are displayed in figure \ref{fig:PrefBranch}: panel (a) shows the value of the wavenumber $k$ along the preferred branch as a function of the temperature $T$,  and panel (b) and (c) show the free energy and entropy of the preferred solutions. 
 We see that at zero temperature, we approach a ground state with a finite wavenumber $k$ that seems to have non-zero entropy. Moreover, in  figure \ref{fig:Thermo_05Tc} we display the the free energy and the averaged stress tensor for $T=0.5\, T_c$ as functions of $k$.  We see that for the preferred branch, at the minimum of the red curve, the green and the blue curves intersect showing that the averaged stress tensor reduces to the one for a isotropic ideal fluid. Moreover, the minimum averaged entropy given by the pink curve is slightly to the right of the preferred $k$ at that temperature. This suggests that the free energy decreaaes faster at that value of $k$ with increasing temperature. This is certainly compatible with the fact that the thermodynamically preferred $k$ changes with $T$.
 
\begin{figure}[h]
\centering
\subfloat{\includegraphics[height=5cm]{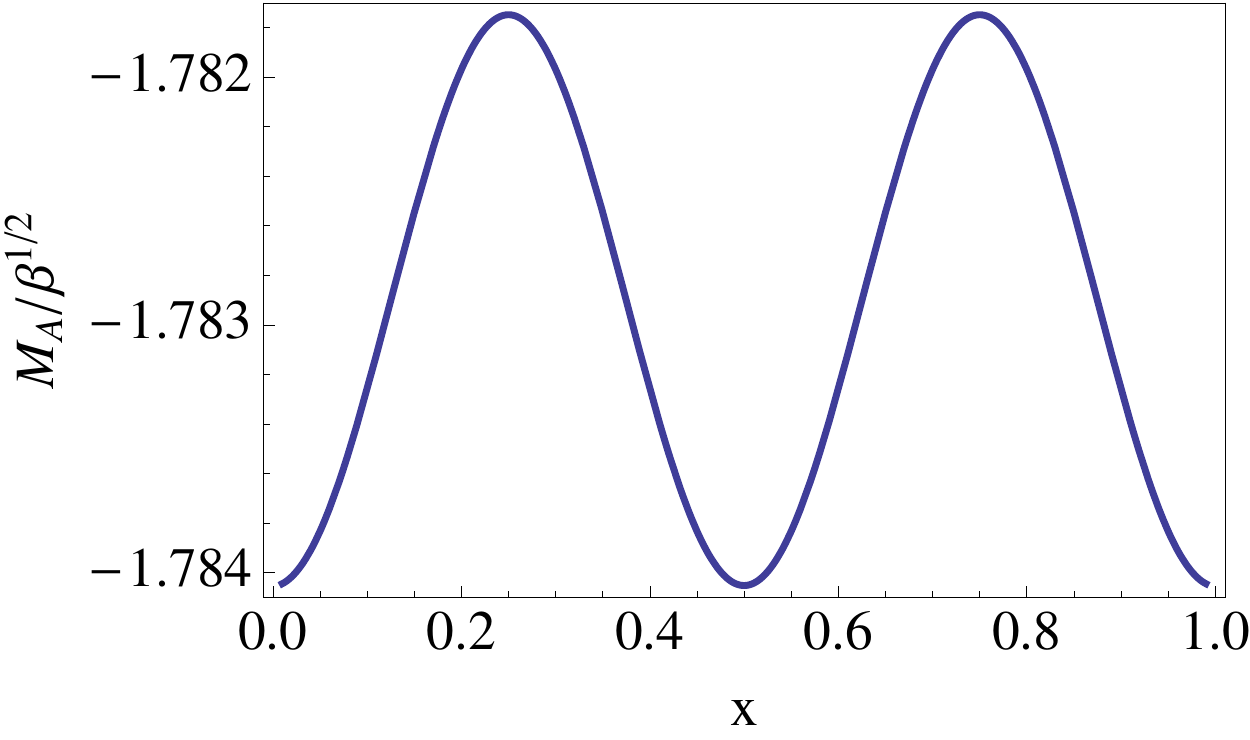}\includegraphics[height=5cm]{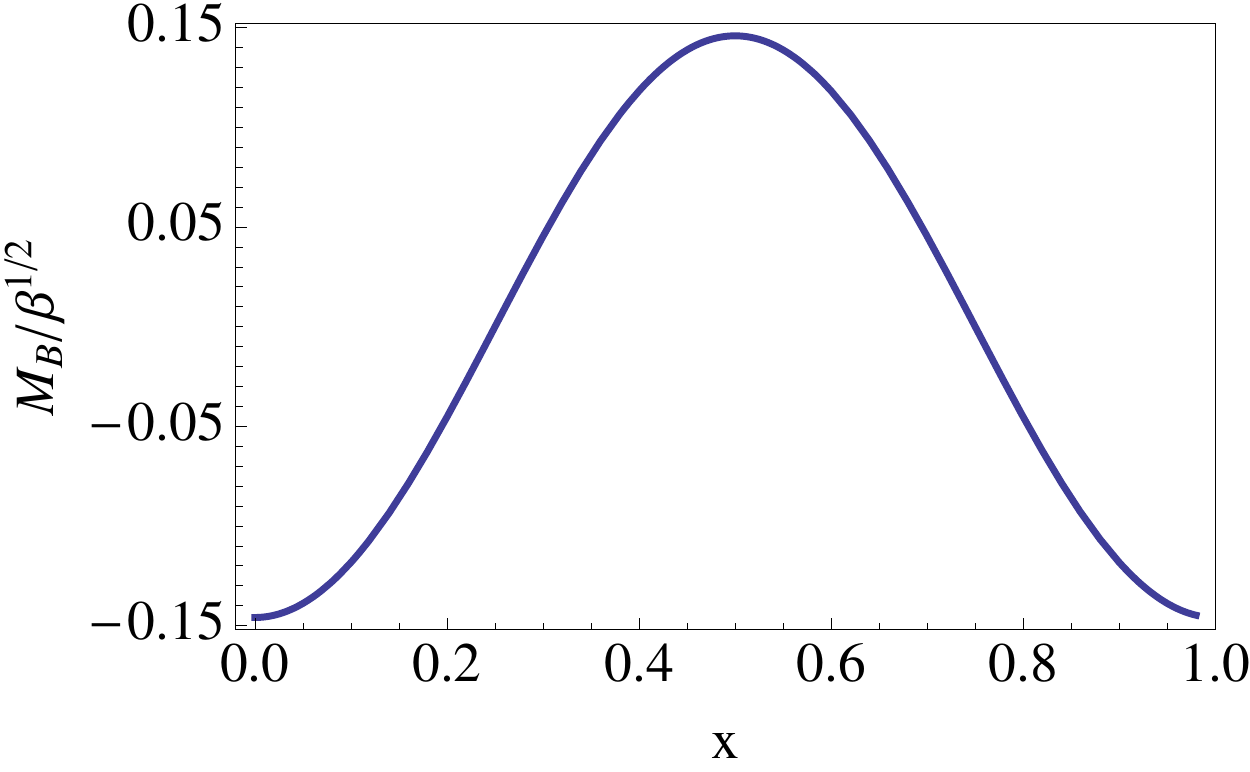}}\\
\subfloat{\includegraphics[height=5cm]{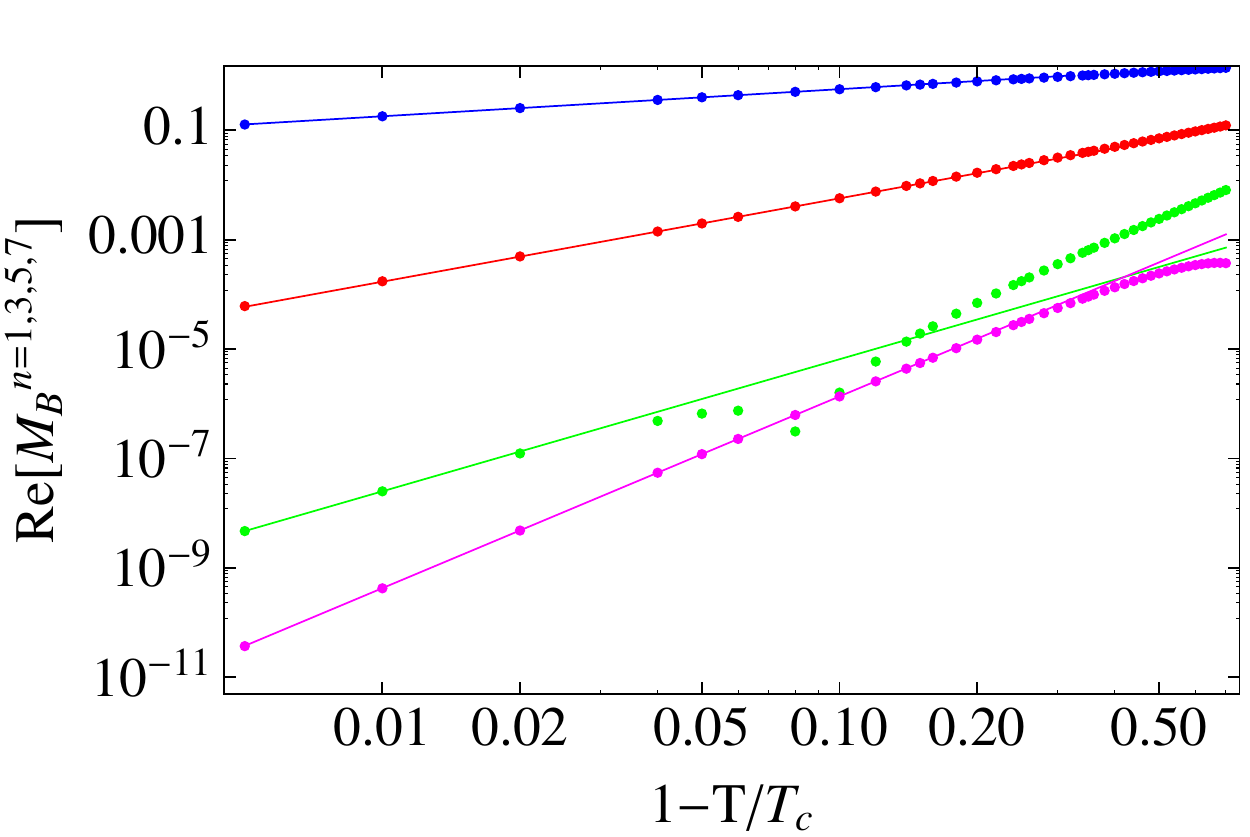}}
\caption{Magnetisation densities of $U(1)_{A}$ (top left) and $U(1)_{B}$ (top right) for $(T,k)=(0.5\, T_c, k_c)$. Field theory direction Fourier modes of the magnetisation density of $U(1)_{B}$ as a function of $T$ and for $k=k_{c}$ (bottom).}
\label{fig:magn}
\end{figure}

 In figure \ref{fig:magn} we plot the magnetisation densities $M_{A}$ and $M_{B}$ for the broken phase solution corresponding to $(T,k)=(0.5\, T_c, k_c)$. As we discussed in section \ref{sec:setup} the leading mode of the instability concerns the magnetisation of $U(1)_{B}$ and figure \ref{fig:magn} agrees with expectation that the modulation of $M_{B}$ is a leading order effect. We can use  equation \eqref{eq:magnetisation_np} to find $M_{A}$ at the same temperature, in the normal phase. For $T=0.5\,T_{c}=0.04\,\beta^{1/2}$ we find that  $M_{A}\approx -1.787$ and a comparison with the top left of figure \ref{fig:magn} demonstrates that the modulation in $M_{A}$ is a higher order effect when compared to that of $M_{B}$.
 
 We have also included a log-log plot of the first four non-trivial Fourier modes $M^{n}_{B}$ of the magnetisation density $M_{B}$ as functions of $1-T/T_{c}$. Using simple perturbative reasoning, one can argue that the behaviour of these should be
 \begin{align}
 M_{B}^{n}(T,k)= M_{B}^{n}(k)\,\left(1-\frac{T}{T_{c}} \right)^{\frac{n}{2}}+\cdots
 \end{align}
 which certainly conforms with our findings shown in the bottom plot of figure \ref{fig:magn}. In particular, we find that due to the $\mathbb{Z}_{2}$ symmetry of our theory, the only non-trivial modes switched on are for $n=1,\,3,\,5,\ldots$.

  \section{Discussion}\label{sec4}

In this paper we have  constructed numerically a two-parameter family of inhomogeneous AdS black brane solutions by backreacting on the instabilities of \cite{Donos:2011qt}, for a particular model. These geometries describe a spatially modulated phase of the dual field theory, held at finite temperature and external magnetic field. Below a critical temperature $T_{c}$ the modulated phase becomes thermodynamically preferred over the unbroken one. We have shown substantial numerical evidence that the transition is second order.

By exploring the thermodynamics of the entire parameter space of these solutions we find that the preferred solutions has a temperature-dependent periodicity for the modulation, in accordance with \cite{Donos:2012gg,Donos:2013woa,Withers:2013kva}. In contrast with previously constructed examples, we find that the periodicity is a monotonically decreasing  function of temperature. Analysing our low temperature solutions, we find indications that the system approaches an inhomogeneous ground state with non-zero entropy. However, this picture could change as soon as we achieve even lower temperatures. It would be interesting to try to construct the corresponding  ground states directly by considering $T=0$.

An obvious question to be addressed in the future is to relax the assumption of translational invariance in the one of the directions and construct the corresponding three-parameter family of solutions. The present work has not excluded the possibility that these solutions could dominate the ensemble. In the context of inhomogeneous phases of field theories at finite temperature and chemical potential, these solutions have been constructed in \cite{Withers:2014sja,Donos:2015eew} and  a competition between phases with modulation in one and two directions was found with the triangular configurations being preferred \cite{Donos:2015eew}.

A more ambitious direction we plan to pursuit in the future is the fate of the instabilities of top-down models discussed in \cite{Donos:2011pn}. We expect a vast number of ground state geometries in these models with the finite temperature phase diagram exhibiting competing orders\footnote{Note that some of the normal phase solutions in these models are known in closed form \cite{Duff:1999gh,Cvetic:1999xp}. More recently these solutions have been used to generate dyonic solutions of theories different from the STU model \cite{Gnecchi:2016auh}.}. Among these ground states, of particular interest are the supersymmetric ones \cite{Almuhairi:2011ws,Donos:2011pn} and in particular the ones which will be modulated \cite{Donos:2011pn}.

 \section*{Acknowledgements}
CP wishes to thank the University of Durham for hospitality and  acknowledges financial support from the EU-COST Action MP1210 ``The String Theory Universe" in the form of a Short Term Scientific Mission (STSM). We acknowledge financial support by grants 2014-SGR-1474, MEC FPA2010-20807-C02-01, MEC FPA2010- 20807-C02-02, CPAN CSD2007-00042 Consolider- Ingenio 2010, and ERC Starting Grant HoloLHC-306605.
 
 \appendix

\section{Boundary conditions}\label{app:boundary_conditions}

To construct the boundary expansion we start by determining the scaling dimensions of the operators in the dual conformal field theory. To do this, we consider the following perturbations around the $AdS_4$ solution
\begin{align}
&Q_{ii}=1+ \delta Q_{ii}(z,x)\,,\quad i=\{t,z,x,y\}\,,\nonumber\\
&Q_{z\,x}=0+ \delta Q_{z\,x}(z,x)\,,\nonumber\\
&h=0+\delta h(z,x),,\nonumber\\
&a_y=0+\delta a_y(z,x)\,,\nonumber\\
&b_y=0+\delta b_y(z,x)\,.
\end{align}
After substituting into the equations of motion, at first order in the perturbation we find that we need to solve a set of second order linear PDEs for the variations. We proceed by looking for solutions where the eight function variation, as a vector, are of the form $\vec{v}\, r^\delta$ where $v$ is a constant vector and $\delta$ is a constant that is related to a scaling dimension in the three-dimensional conformal field theory dual to the $AdS_4$ solution. The system of equations then takes the form $M\cdot\,v = 0$ where $M$ is an $8 \times8$ matrix that depends on $\delta$. Demanding that non-trivial values of $v$ exist implies that $det\,M = 0$ and this specifies the possible values of $\delta$. The solutions come in 8 pairs. Apart from the usual modes,
\begin{align}
\delta_1=\{0,1\},\quad \delta_2=\{0,1\},\quad \delta_3=\{0,1\},\nonumber\\
\delta_4=\{0,3\},\quad \delta_5=\{0,3\},\quad \delta_6=\{-3,2\},
\end{align}
 corresponding to the scalar, the 2 gauge fields and the metric respectively, we also obtain two modes with scaling dimensions
\begin{equation}
\delta^\pm_{7,8}=\frac{1}{2}(3\pm\sqrt{33})\,,
\end{equation}
that appear in the variations of the metric functions. These modes appear only in the Einstein-De Truck equations and are manifestations of the dynamical gauge fixing procedure. 

 We then proceed in constructing the actual expansion
  \begin{align}
  \label{UVexp}
& Q_{tt}=1-\frac{1}{4} \phi_1^2(x)\, z^2+Q_{tt}^{(3)}(x)z^3+Q_{tt}^{(4)}(x)z^4+g_1(x) z^{(3+\sqrt{33})/2}+\mathcal{O}(z^5\,log(z))\,,\nonumber\\
& Q_{zz}=1+Q_{tt}^{(4)}(x)z^4+g_2(x) z^{(3+\sqrt{33})/2}+\mathcal{O}(z^5\,log(z))\,,\nonumber\\
 &Q_{xx}=1-\frac{1}{4} \phi_1^2(x)\, z^2+Q_{xx}^{(3)}(x)z^3+Q_{xx}^{(4)}(x)z^4+g_1(x) z^{(3+\sqrt{33})/2}+\mathcal{O}(z^5\,log(z))\,,\nonumber\\
 &Q_{yy}=1-\frac{1}{4} \phi_1^2(x)\, z^2+Q_{yy}^{(3)}(x)z^3+Q_{yy}^{(4)}(x)z^4+g_1(x) z^{(3+\sqrt{33})/2}+\mathcal{O}(z^5\,log(z))\,,\nonumber\\
 &Q_{zx}=\frac{z_h^2}{8 L} \phi_1(x)\, \phi_1^{\prime}(x)\, z+Q_{zx}^{(2)}(x) z^2-\frac{z_{h}^{2}\,z^2\, \ln(z)}{5 L}\left(Q_{xx}^{(3)'}(x)+L z_h^2\,\beta \,j_a(x)\right)+\mathcal{O}(z^3\,log(z))\,,\nonumber\\
 &h=\phi_1(x)+\mathcal{O}(z^2)\,,\nonumber\\
 &a_y=j_a(x) z+\mathcal{O}(z^2)\,,\nonumber\\
 &b_y=j_b(x) z+\mathcal{O}(z^2)\,,
 \end{align}
where the functions $Q_{\mu\mu}^{(4)}(x)$ are fixed in terms of $\phi_{1}$, $j_{a}$ and $j_{b}$. The expansion \eqref{UVexp} bares a lot of similarities with the expansions that appeared in \cite{Donos:2014yya,Donos:2015eew} in related topics. As we said above, we choose the operator dual to the scalar field to have scaling dimension $\Delta=1$, which gives $\phi_1$ the interpretation of the vacuum expectation value and $\phi_2$ is the source. We want the breaking of translation invariance to be spontaneous and thus, we require the sources of all the operator, except of the background magnetic field, to vanish. For this reason, we impose $\phi_2=\mu_a=\mu_b=0$. The expansion \eqref{UVexp}, is then specified in terms of nine  coefficients $\{Q_{tt}^{(3)},Q_{xx}^{(3)},Q_{yy}^{(3)},Q_{zx}^{(2)},\phi_1(x),j_a(x) ,j_b(x), g_1(x), g_2(x)\}$ that will be fixed by solving the PDEs  subject to the constraint 
 \begin{align}
&Q_{tt}^{(3)}(x)+Q_{xx}^{(3)}(x)+Q_{yy}^{(3)}(x)=0 \,,
 \end{align}
which reflects the fact that the energy-momentum tensor of the dual field theory is traceless, $\langle T^{\mu}{}_\mu\rangle=0$. Demanding that $\xi^2=0$ poses extra constraints, namely that $g_2=\frac{-1}{2}(3+\sqrt{33})\,g_1$ and the Ward identity given by $Q_{xx}^{(3)'}(x)=\frac{2}{3}\,L z_h^2\,\beta \,j_a(x)$. Note that in the solutions we will construct in the remaining of this paper the coefficients $g_1, g_2$ are non-trivial, but they can be removed by a gauge transformation. However, even though they are pure gauge, these terms do leave an imprint on our numerics as, together with the logarithmic terms, they compromise the convergence rate.

 \section{Numerical tests}\label{app:Numerical}
In this appendix we discuss two tests we performed in order to check the quality of our numerics. The first one is to check the first law of thermodynamics, while the second one involves the convergence of $\xi^2$. 

\begin{figure}[h]
\centering
\subfloat{\includegraphics[width=7cm]{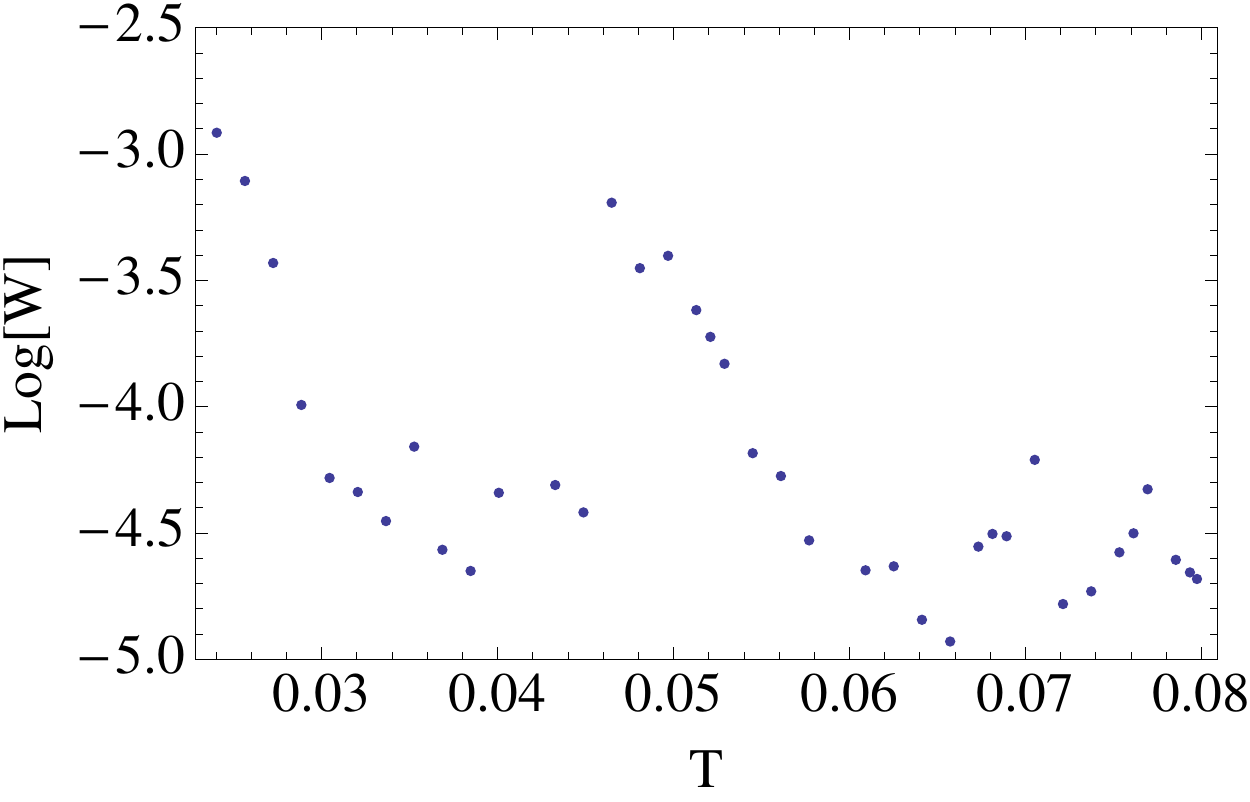}\quad\includegraphics[width=7cm]{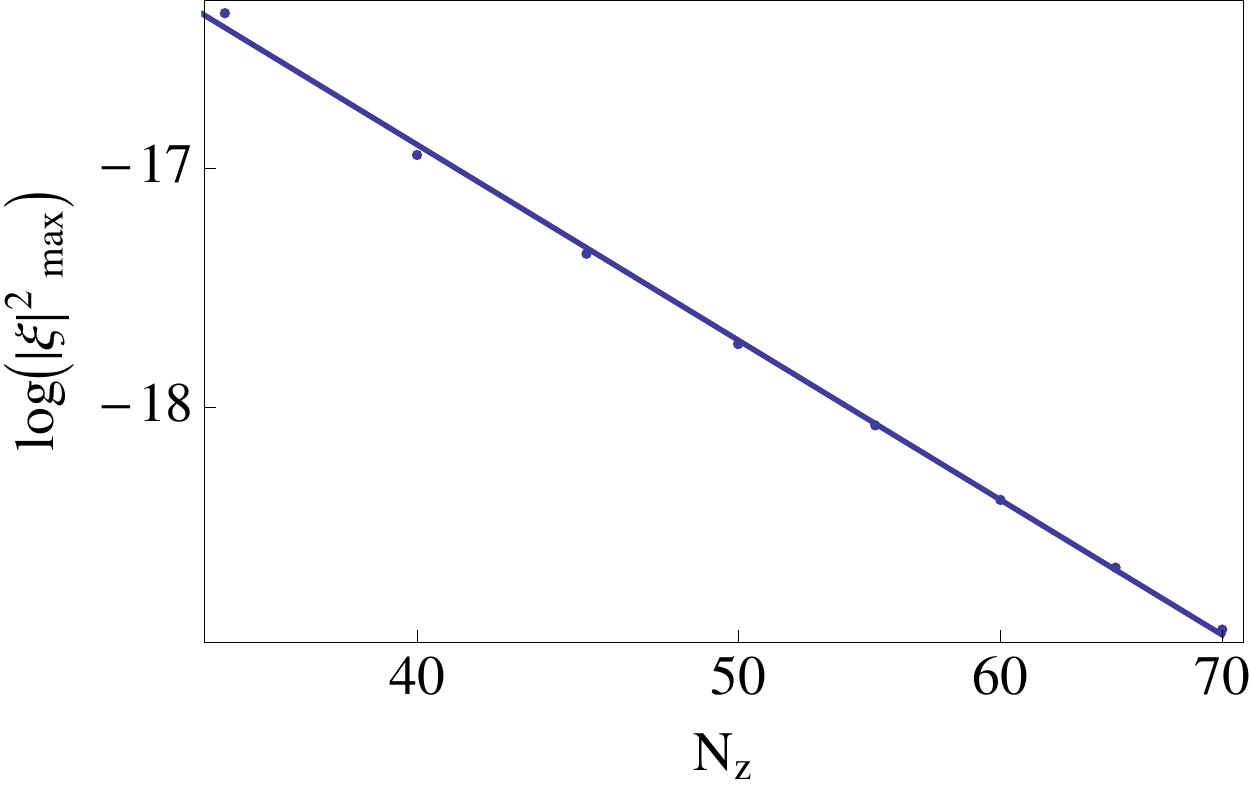}}
\caption{(a)Test of the first law of thermodynamics on the branch with $k=k_c$ for various temperatures. (b) Convergence of $|\xi|^2_{max}$ as function of $N_z$ for $T=0.5 \,T_c$, $k=k_c$ and $N_x=50$. Both figures correspond to model \eqref{conds2}.}
\label{fig:NumTest}
\end{figure}
   
From the differential form of the first law \eqref{eq:first_law}  we conclude that along a fixed $k$ branch of solutions the quantity $W=\bar{S}+\partial_{T}\bar{w}$ must vanish. In particular, using the values of $w$ for equally spaced values of temperature $T$ to construct an interpolating polynomial of degree 4 and evaluate $\partial_{T}\bar{w}$. Indeed, in figure \ref{fig:NumTest}(a) we see that, for model \eqref{conds2}, this condition is met along the set of solutions with $k=k_c$, where the temperature step between successive points used for the interpolation was $\Delta T \sim10^{-3}\,\beta^{1/2}$. The change in the behaviour of $W$ at $T=0.05\,\beta^{1/2}$ is justified by the fact the we changed our grid resolution at that point from $N_{z}=40$ to $N_{z}=60$ as we lowered the temperature.

One should also illustrate numerical convergence as the number of grid points is varied. For the DeTurck quantity, $\xi$, defined in section \ref{sec:setup}, we compute its maximum value on the grid, denoted by $|\xi|^2_{max}$. In figure \ref{fig:NumTest}(b) we present the convergence of $|\xi|^2_{max}$ with the number of grid points in the $z$ coordinate, $N_z$, for a representative solution with temperature  $T =0.5\, T_c$ and momenta $k= k_c$ in the model \eqref{conds2}.  The convergence to zero is power law, $N_z^{-8.4}$, and not exponential as one would naively expect for spectral methods. This is due to the appearance of non-analytic terms with leading power $z^{8.7}$ in the UV boundary expansion of $\xi^{2}$. 

\newpage
\bibliographystyle{utphys}
\bibliography{magneticstripes}{}
\end{document}